\documentclass [12pt,a4paper]{article}

\usepackage{graphicx}
\usepackage{epsfig}
\usepackage{amsfonts}
\usepackage{amssymb}
\usepackage{amsmath}
\newcommand{\D}[2]{\frac{\partial #2}{\partial #1}}

\renewcommand{\vec}[1]{\mbox{\boldmath$#1$}}

\begin{document}

\title{Model Reduction Applied to Square to Rectangular
Martensitic Transformations Using Proper Orthogonal Decomposition }

\author{L. X.  Wang $^1$
  \thanks{Corresponding Address:Tel.:+45 6550 1686, Fax:+45 6550 1660,
  E-mail:wanglinxiang$@$mci.sdu.dk}
\quad and \quad Roderick
V.N. Melnik $^2$   \\
$^1$  MCI, Faculty of Science and Engineering, \\
 University of Southern Denmark,\\
 Sonderborg, DK-6400, Denmark \\
 $^2$ Mathematical Modelling \& Computational Sciences, \\
  Wilfrid Laurier University, \\
 75 University Avenue West, \\
 Waterloo, ON, Canada, N2L 3C5 \\
 }

\date{}

\maketitle


\begin{abstract}Model reduction using the proper orthogonal decomposition
  (POD) method
is applied to the dynamics of ferroelastic patches to study the first order
square to rectangular phase  transformations.
 Governing equations for the system dynamics are constructed
 by using the Landau-Ginzburg theory and  are solved numerically. By using the
 POD method, a set of empirical orthogonal
basis functions is first constructed, then the system is projected onto the subspace spanned by a small set of
basis functions determined by the associated singular values. The performance of the low
dimensional model is verified by simulating nonlinear thermo-mechanical waves
and square to rectangular transformations in a ferroelastic patch. Comparison
between numerical results obtained from  the original PDE model and the low
dimensional one is carried out.

 \end{abstract}


\section{Introduction}

\noindent  The first order martensite transformation is a common transformation
in many ferroelastic, ferromagnetic, and ferroelectric materials that holds the
key to their  specific properties.  For many new engineering applications of such
materials,  a better
understanding of the mechanism of the transformation becomes decisively
important. As a result, the martensite transformations observed in smart
materials  have attracted substantial attention from  engineers, mathematicians,
physicists, and control theorists.   These materials possess unique properties
of being able to sense, actuate, and inherently respond to  external
stimuli  (\cite{Sood2001} and references therein). Much attention  is devoted  to control and design
optimization of these materials where mathematical modelling tools  become ubiquitous.
However, even for the systems governed by  linear Partial Differential Equations (PDE), these
issues are far from trivial \cite{Rowley2004,Teman1998}.  
At the same time,  mathematical models  describing the dynamic behavior of
most smart materials are intrinsically nonlinear, coupling different physical
fields by conservation law equations and constitutive relations (e.g.,
\cite{Lookman2003,Melnik2001,Wang2004}). To bridge the gap between the nonlinear PDEs such as those arising in the
description of the smart material systems  and the low dimensional models
suitable for control and optimization in realistic engineering applications, 
it is natural to attempt to approximate the dynamic behavior of PDEs by low
dimensional models.  Such low dimensional models can be obtained from the
original systems by model
reduction and the POD provides an efficient tool for doing so as soon as a collection of system states is
available. This idea has been successfully applied to many active control
problems involving fluid flows (see
\cite{Holmes1997,Rowley2004} and references therein). Recently, we applied
this idea to the analysis of phase transformations in the 1D case \cite{Wang2005}. In its
essence, the  POD methodology is close to the principal
component analysis, techniques based on the singular value decomposition, and
Karhunen-Loeve decomposition \cite{Holmes1997,Rowley2004}.

In this paper, we develop a low  dimensional model for the dynamics of
ferroelastic patches exhibiting square to rectangular phase
transformations to study  2D nonlinear thermomechanical waves. First, the
original system of PDEs is analyzed numerically with  Chebyshev's collocation method.
Then, the basis functions are extracted from the obtained numerical results using the
POD. Finally, the low dimensional model is  constructed using the Galerkin
projection method.  The dynamic behavior of ferroelastic patches is
simulated by the empirical low dimensional model, and its performance is
evaluated by comparing the numerical results with those from  the original PDE
model.


\section{PDE Model and Its Numerical Solution}

Square to rectangular transformations are regarded as a 2D analog of
the cubic to tetragonal or  tetragonal to orthorhombic transformations observed
in the general 3D case (see, e.g., \cite{Lookman2003} and reference
therein).  At mesoscale, there is a
high temperature phase  of greater symmetry (austenite, represented schematically as square), and two lower
temperature phases (martensite variants, represented schematically as rectangles). The transformation from
austenite to martensite is  not unique due to the fact that there are
multiple equivalent orientations of the lower symmetry lattice with respect to the
parent lattice.The transformation can be induced by either mechanical or thermal
loadings (or their combination), but  only mechanically induced transformation will be modelled here.

The governing equations describing the macroscopic dynamics of the ferroelastic
structure exhibiting the above mentioned transformation can be formulated on the
basis of conservation laws. Following \cite{Wang2004}, we consider the
following coupled system of PDEs based on conservation of momentum and energy:
\begin{equation}\label{DAESys}\begin{array}{l}  \displaystyle
\rho \frac{\partial^2 u_{1}}{\partial t^2}  = \frac{\partial \sigma_{11}}{\partial x}+
\frac{\partial \sigma_{12}}{\partial y}+ f_{x}, \quad
\rho  \frac{\partial^2 u_{2}}{\partial t^2} =  \frac{\partial \sigma_{12}}{\partial x}+
\frac{\partial \sigma_{22}}{\partial y} + f_{y},
\\[10pt]  \displaystyle
c_v \frac{\partial \theta}{\partial t} = k \left (\frac{\partial^2
\theta}{\partial x^2} + \frac{\partial^2 \theta}{\partial y^2} \right )
+  a_2 \theta e_2 \frac{\partial e_2}{\partial t} + g ,
  \\[10pt]  \displaystyle
  \sigma_{11} =  \frac{\sqrt{2}}{2} (
a_1 e_1 + a_2 (\theta - \theta_0) e_2 - a_4 e_2 ^3 +  a_6 e_2 ^5 )
     + \frac{d_2}{2} \nabla_x^2 e_2   ,
   \\[10pt] \displaystyle
\sigma_{12} =   \frac{1}{2}   a_3 e_3  = \sigma_{21} ,
   \\[10pt] \displaystyle
\sigma_{22} = \frac{\sqrt{2}}{2}  (
a_1 e_1 -  a_2 (\theta - \theta_0) e_2 + a_4 e_2 ^3 -  a_6 e_2 ^5
     + \frac{d_2}{2} \nabla_y^2 e_2 ),
\end{array}
\end{equation}

\noindent where $c_{v}$ is the specific heat constant, \textit{$\theta$}$_{0}$ is
the reference temperature for the transformation,  $\nabla$ is the gradient
operator, $a_1~a_2,~a_3, ~a_4, a_6$,  and  $d_2$ are the material-specific
coefficients, and $e_{1}$, $e_{2}$, $e_{3}$ are dilatational, deviatoric, and
shear components of the strains, respectively.  $f_x, f_y$
and $g$ are mechanical and thermal loadings, respectively.

In the above model, $e_2$ is chosen as the only
order parameter to characterize differences in  phases. Strain components are
defined as follows:
\begin{equation}\label{Strain}
\begin{array}{l} \displaystyle
e_{1}=\left(\eta_{11}+\eta_{22}\right)/\sqrt{2},
     \\[10pt] \displaystyle
e_{2}=\left(\eta_{11}-\eta_{22}\right)/\sqrt{2},
        \\[10pt] \displaystyle
e_{3}=\left(\eta_{12}+\eta_{21}\right)/2.
\end{array}
\end{equation}

\noindent where the Cauchy-Lagrangian strain tensor $\vec{\eta}$ is given by its
components as follows(with the repeated-index convention used):

 \begin{equation}
\eta_{ij}\left(\textbf{x},t\right)=\left(\frac{\partial
u_{i}\left(\textrm{\textbf{x}},t\right)}{\partial x_{j}}+\frac{\partial
u_{j}\left(\textbf{x},t\right)}{\partial x_{i}}\right)/2,\kern0pt
\label{LagStrain}
\end{equation}

In the above formulation,  stress components are kept as independent variables
to be solved for. This approach is similar to the DAE approach first proposed
in the context of shape memory alloys materials modelling in \cite{Melnik2001}.

\subsection{Chebyshev Collocation Methodology}

The system formulated in the previous section is solved here numerically by
the method of lines.  First, we discretize the system  spatially, converting
it into a system of differential-algebraic equations, and then we apply a
backward differential formula to integrate the resulting system in time.

For the spatial discretization, the Chebyshev pseudospectral approximation is
employed on a set of 2D Chebyshev points $(x_i, y_j ) \in [-1,1] \times
[-1,1]$  \cite{Alfio2000}:
 \begin{equation}
\label{chebpoints}  x_i  =  \cos(\frac{\pi i}{N}) ,  y_j  =  \cos(\frac{\pi j}{N})
,\quad i,j=0,1,\dots,N,
\end{equation}

\noindent where $N+1$ is the number of nodes. Using the discretization, all the
unknown distributions in the patch can be approximated by the following
linear combination:
 \begin{equation}
  \label{chebapprox}
   f(x,y) = \sum_{i=0}^{N} \sum_{j=0}^{N} f_{i,j}
   \phi_i(x)\phi_j(y),
\end{equation}

\noindent where $f(x,y)$ could be either stresses, displacements, velocities, or
temperature; $f_{i,j}$ is the function value at $(x_i,y_j)$; $\phi_i(x)$ and
$\phi_j(y)$  are the $i^{th}$ and $j^{th}$ Lagrange interpolating polynomials in
$x$ and $y$ directions, respectively.

Because Eq.(\ref{chebapprox}) is just a linear combination of the interpolating
polynomials, the derivatives of the functions $\partial f(x,y)/ \partial x$ and
$\partial f(x,y)/\partial y$ can be easily obtained by calculating $\partial
\phi_i(x)/ \partial x$ and $\partial\phi_j(y)/ \partial y$. Following the idea
given in\cite{Trefethen2000}, all the differential operators in Eq.(\ref{DAESys})
can be written as a matrix form:
\begin{equation}
\label{chebdiff}
\frac{\partial f(x,y)}{\partial x} \Big
\arrowvert_{x_i,y_j} =  \vec D_x \vec F , \quad
\frac{\partial f(x,y)}{\partial y} \Big \arrowvert_{x_i,y_j}
  =  \vec D_y \vec F,
\end{equation}

\noindent and similarly for the higher order differentiation operators. The
differentiation matrices $ \vec D_x$, $\vec D_y$, $\vec D_{xx}$, and $\vec D_{yy}$
can be calculated using the approximation given by Eq.(\ref{chebapprox}).

By approximating all the differential operators in the governing 
Eq.(\ref{DAESys}), the system can be discretized and converted into a set of
nonlinear ODEs and algebraic equations. The boundary conditions of the system can
be easily enforced  \cite{Trefethen2000,
Alfio2000}. The system
after discretization finally takes the following form:
 \begin{equation}
    \label{DAEDiscret}
      \vec M \frac{ d \vec X }{dt} + \vec G \left(
 t, \vec X, \vec U \right)  = \bf{0} ,
\end{equation}

\noindent where $\vec X$ is a $ 5(N+1)$ vector collecting all the unknowns
we are seeking for, the matrix $\vec M= {\rm diag}(a_1,a_2,...,a_{5(N+1)})$
is a $5(N+1) \times 5(N+1)$ matrix having entries ``one'' for all
the differential equations associated with ODEs  and ``zero'' for
all algebraic equations.

The above DAEs system is a stiff system and must be solved by an implicit
algorithm. Here, the second order backward differentiation formula method
is employed for this purpose. By discretizing the time derivative
using the second order approximation, the DAE system can be converted into the
following algebraic system at each time level:
\begin{equation} \bf M \left(
\frac{3}{2} \bf X^n - 2 \bf X^{n - 1} + \frac{1}{2} \bf X^{n - 2} \right) + \Delta
t \bf N \left( t_n, \bf X^n,V(t_n) \right) =0 ,
\end{equation}

\noindent where $n$ denotes the current computational time layer. For each
computational time layer, iterations are carried out with the bi-conjugated
gradient iteration method for computing $\bf X^n$ having $\bf X^{n-1}$
and $\bf X^{n-2}$. Starting from the initial value, the vector of unknowns ${\bf
X}$ is sought for  all specified time instances employing this algorithm.

We note that to deal with strong (cubic and quintic) nonlinearities in the
order parameter, a smoothing procedure similar to that proposed in
\cite{Niezgodka1991} has been employed during the iteration process. In
particular, we have used the following expansions:
\begin{equation}\label{eq4-4}
\begin{array}{l}
\displaystyle
y^{3}=\frac{1}{4}\sum\limits_{i=0}^{3}y_n^{i}y_{n-1}^{3-i}, \quad
y^{5}=\frac{1}{6}\sum\limits_{i=0}^{5}y_n^{i}y_{n-1}^{5-i},
\end{array}
\end{equation}

\noindent where $y$ stands for $e_2$ here, and $n$ time layer.


\section{Construction of Low Dimensional Models}

\subsection{Proper Orthogonal Decomposition}

One of the most effective ways to construct a low dimensional model for approximation to a
 dynamical system is to find a set of optimal basis functions by which the dynamical system
can be approximated best at any time. In what follows, we apply the POD method
 to construct an  empirical basis for the system
dynamics describing 2D phase transformations.

For the dynamical system given by Eq.(\ref{DAESys}), the system state at any time
can be fully characterized by the displacements,  velocity, and
temperature distributions, while displacements and velocities can be
represented by using the same basis functions. Here, we only need  to construct
spatial basis functions for displacement and temperature distributions. Let us
assume that $\mathcal U(x,t)$, which describes the system dynamics,  belongs to an inner product
(Hilbert) space $\mathcal H $. The POD is concerned with the possibility to find a
set of orthonormal spatial basis functions $\phi = \{ \phi_j(x), j=1,\ldots,P \}$
in $\mathcal H$ which are optimal in the sense that the $P$ dimensional
approximation
  \begin{equation}    \label{podeq-2}   \mathcal{U}_P(x,t) =
\sum_{i=1}^{P}a_i(t) \phi_i(x)
 \end{equation}

\noindent gives the best (in the least
square sense) approximation to the function $\mathcal U(x,t)$ among
all those $P$ dimensional approximations in $\mathcal H$  \cite{Holmes1997,Rathinam2003}. Here $a_i$ are the general Fourier
coefficients associated with $\{\phi_i\}$. They are functions of time.
Mathematically, the idea of POD is to choose the basis
functions $\{\phi_k \}$ to maximize the mean projection of the function  $\mathcal
U(x,t)$ on  $\{\phi_k\}$
\begin{equation}
  \label{podeq-3}
\max_{\phi \in L_2(\Omega)} \frac{ E \left( | \langle \mathcal{U},\phi
 \rangle |^2 \right )}  {\| \phi \|^2 }
\end{equation}

\noindent where $E({\bf \cdot})$ denotes the mean value functional, $\vert~
{\bf \cdot}~\vert$ is the modulus, $\langle~ {\bf \cdot}~\rangle$ is the inner
product, and $\Vert ~ {\bf \cdot} ~\Vert $ is the
$L^2$ norm \cite{Holmes1997,Rathinam2003}.

For convenience of the projection operation, that will be employed later, the
orthonormality requirement is enforced on the basis functions we are
looking for:
 \begin{equation}  \label{podeq-5}
 \langle \phi_i, \phi_j \rangle = \left \{
    \begin{array}{ll}
        1 & \quad  \textrm{if} \quad   i=j,      \\
        0 &  \quad  \textrm{if} \quad  i \neq  j.   \\
    \end{array}  \right.
\end{equation}

\noindent The general Fourier coefficients in this case can be calculated
simply as:
\begin{equation}
    \label{podeq-6}
   a_i(t) = \langle \mathcal U(x,t), \phi_i \rangle.
\end{equation}

Taking into account the above requirement on the basis functions,
the maximization problem can be reformulated in terms of calculus of variation,
with a functional for the constrained variational problem
\cite{Holmes1997,Rathinam2003}:

\begin{equation}
  \label{variation}
J[\phi] =    E \left( | \langle \mathcal{U},\phi
 \rangle |^2 \right ) - \lambda (\| \phi \|^2 -1 ).
\end{equation}

\noindent A necessary condition for extrema is that the first order variation of
the functional with respect to all variations $\phi + \delta \psi$ should be zero:

 \begin{equation}
 \D{\delta}{}J[\phi+\delta \psi] \arrowvert_{\delta =0} = 0,
 \end{equation}

\noindent which finally leads  to the following eigenvalue problem
\cite{Holmes1997,Rathinam2003}:
\begin{equation}
   \label{podeq-4}
\int_{\Omega} E(\mathcal{U}(x)\mathcal{U}(x'))\phi(x')dx' = \lambda \phi(x).
\end{equation}

\noindent whose kernel $\mathcal K = E(\mathcal{U}(x)\mathcal{U}(x'))$ is the
auto-covariance function of the two points $x$ and $x'$.

For the current problem, the dynamics of the system can be sampled at any time
instances we need, by using the numerical algorithm mentioned
above for Eq.(\ref{DAESys}).  Let us denote by $U^{i}$ the system state at the
$i^{th}$ time instance, called the $i^{th}$ snapshot. In discrete form, each
snapshot can be written as a column vector with $M$ entries, where $M$ is the
number of nodes for spatial discretization.

To construct the orthonormal basis, all the snapshots are collected in one
matrix $ U = \{ U^{i}, i=1,\ldots,N \}$.  If one put each snapshot as one column
in the collection matrix,  the collection $U$ will be a matrix of $ M \times N$.
The orthonormal basis vectors for the given collection $U$ can be calculated by
the singular value decomposition as follows:

 \begin{equation}
   \label{podeq-7}
 U = L S R^T
\end{equation}

\noindent where $L$ is $M\times M$ orthonormal matrix, $R$ is
a $N\times N$ orthonormal matrix and the superscripts $T$ indicates matrix
transpose. $S$ is a $M\times N$ matrix with all elements zero except along the
diagonal, those non-zero elements are arranged in a decreasing way along the
diagonal. They are the singular values of $U$, emphasizing the 
relative importance of its associated eigenvectors in $L$ and $R$.

Let $SR^T=Q^T$ in the singular value decomposition, then $U = LQ^T$. Let
$\phi_k$ be the $k^{th}$ column of $L$ and $a_k$ be the $k^{th}$ row of
$Q$, then the matrix $U$'s singular value decomposition can be rewritten as

\begin{equation}
  \label{podeq-8}
  U = \sum_{k=1}^{m}a_k \phi_k
\end{equation}

\noindent where $ m = min(N,M)$ is the rank of the collection matrix $U$. This
approximation is what we are looking for in Eq.(\ref{podeq-2}). The lower
dimensional approximation of the matrix $U$ can be easily obtained by 
keeping the first few largest singular values and their associated eigenvectors
in $L$ and $Q$. The number of eigenvectors should be determined by
compromising between the dimension number of the resultant system and the
approximation accuracy \cite{Rathinam2003,Holmes1997}.

\subsection{Galerkin Projection}

It has already been mentioned that  Eq.(\ref{DAESys})  can be converted into a system of
DAEs like  Eq.(\ref{DAEDiscret}) by discretization. It can  also be converted into
a system of ODEs if we substitute all the constitutive relations into
Eq.(\ref{DAESys}). The latter approach is applied here for the construction of
 the low dimensional model. For
convenience, Eq.(\ref{DAESys}) is converted into a dynamical system representation as follows:
  \begin{equation}
   \label{Dynamics}
  \frac{d}{dt}  \vec X =  \vec Q \left( t, \vec X ,  \vec  H  \right)  ,
\end{equation}

\noindent where $\vec Q$ is the collection of functions given by the spatial
discretization of  Eq.(\ref{DAESys}), while $\vec H$ is the collection of inputs after
conversion.  Vector  $\vec x$ collects all unknowns including displacements in
the $x$  and $y$ directions, velocities in the $x$ and  $y$ directions, and
temperatures in the collocation nodes.

Having obtained the optimal basis functions for the dynamical system, its lower
dimensional approximation can be obtained by projecting the full system
orthogonally onto the subspace spanned by the chosen basis functions. It is
known that the model
reduction can be achieved due to the fact that much smaller number of basis
functions are needed to approximate the full system, when the chosen basis
functions are optimal.

Let us denote the projection operator as $P_r$ and $P_r^T$ its
transpose. Then,  a low
dimensional system on the chosen subspace $S$ can be constructed by using
the following rule: 

\noindent
For any point  $\vec Z \in S $, compute the vector field $\vec
Q(t, \vec H, P_r^T \vec Z )$ and take the projection $P_rQ(t,  \vec H,
P_r^T\vec Z) $ on to the subspace $S$. It should be equal  to
$\vec{\dot Z}$.  The result can be formulated as:
 \begin{equation}   \label{LDSys}
  \frac{d}{dt}  \vec Z =  P_r \vec Q \left( t, P_r^T\vec Z ,  \vec  H  \right).
\end{equation}

The approximation will introduce an error defined as follows:
    \begin{equation}
        r  =  \vec X  - P_r^T \vec Z.
      \end{equation}

To achieve the best approximation with the given basis, the error function of the
approximation is required to be orthogonal to all the basis functions at any time:
\begin{equation}
\label{galerkin-2}
(r,\phi_k) = \int_{\Omega} r(x)\phi(x)dx = 0.
\end{equation}

\noindent It is easy to check that the low dimensional system
given by Eq.(\ref{LDSys}) satisfies this condition, if the orthogonality of the
basis functions is used.


\section{Numerical Results}

To demonstrate  the performance of the proposed methodology,   numerical results from
the low dimensional model for the dynamics of a ferroelastic patch are compared with its
counterparts from the PDE model. The dynamical behavior of
a $\textrm{Au}_{23}\textrm{Cu}_{30}\textrm{Zn}_{47}$ patch with size of
$[0,1]\times [0,1] cm^2$ is analyzed  numerically.
All the physical parameters for this specific material are available in the
literature (e.g., \cite{Melnik2002,Wang2004}).
It has been shown that the first
order square to rectangular transformation can be mechanically induced in this
material \cite{Melnik2002,Melnik2001,Wang2004}. It has also been shown that the
nonlinearity is very strong, and the nonlinear coupling between the mechanical and
thermal fields complicates the problem further.

The strategy here is as follows:

\noindent
We first perform the numerical simulation
using Eq.(\ref{DAESys}) with a representative mechanical load. As a result, the
collection matrix can be constructed and the empirical eigenfunctions can be
extracted. Then, we simulate the dynamics using Eq.(\ref{LDSys}) with
loads that differ from the previous loading.  By comparing the numerical results from
Eq.(\ref{LDSys}) with those from Eq.(\ref{DAESys}), we  assess the
performance of the low dimensional model.

The initial conditions for all the simulations reported below are set the
same. They are
$\theta(x,0) = 250^oK$ and $u_1=u_2=v_1=v_2=0$.  Boundary conditions for
Eq.(\ref{DAESys}) are taken as pinned-end mechanically and insulated thermally.
The distributed mechanical loading for collecting snapshots is the following
(for one period, $g/\left(ms^{2}cm^2\right)$)
    \begin{displaymath}
      f_x=f_y= \begin{cases} 7000 t/2  &  0 \leq t \leq  2    \cr
                 7000 (4- t)/2  &  2 \leq t \leq  4    \cr
	             0          &  4 \leq t \leq  6    \cr
	         7000 (6 -t)/2 &   6 \leq t \leq  8    \cr
	         7000 (t-10)/2 &   8 \leq t \leq  10    \cr
		     0         &  10 \leq t \leq 12.    \cr
	  \end{cases}
    \end{displaymath}

\begin{figure}          \begin{center}
 \includegraphics[height=8cm, width=10cm]{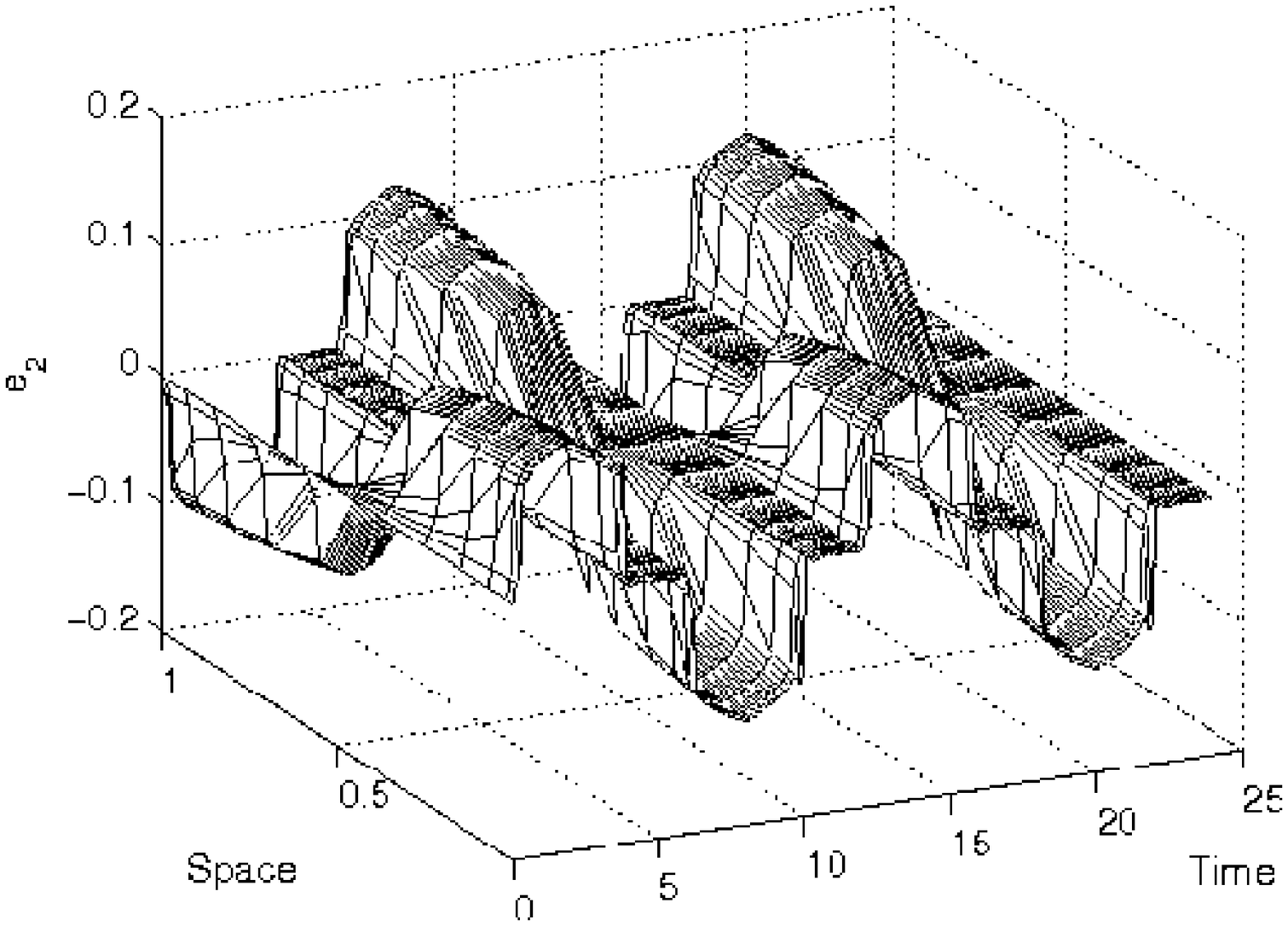} \\
 \includegraphics[height=8cm, width=10cm]{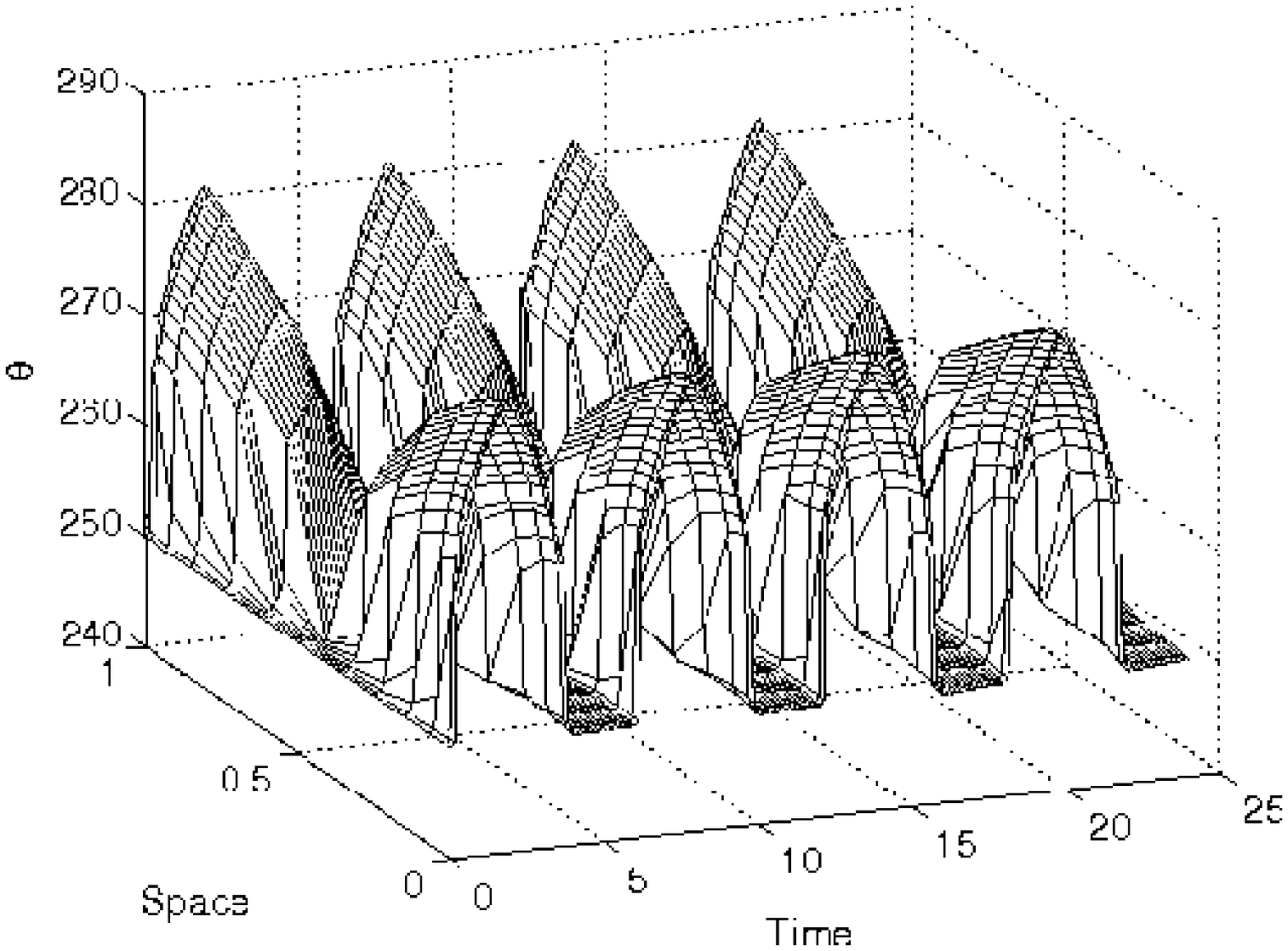}
\caption{Simulated evolution of strain and temperature on the central horizontal line
in the patch involving phase transformation, using PDE model.  (top) Strain evolution, (bottom)
Temperature evolution. }
 \label{PDEResult}
  \end{center}         \end{figure}

There are 13 nodes, each used for $u_1, u_2, v_1, v_2$, and $\theta$
discretization. Two periods of loadings are performed ($t\in [0,24] ms$) and
totally 201 evenly distributed snapshots are sampled. The collection matrix for each field variable
then can be constructed by collecting their values on each sampling instance. To demonstrate the
 phase transformation, the order parameter and temperature on the central horizontal
line are plotted versus time in Fig.\ref{PDEResult}.

\begin{figure}          \begin{center}
 \includegraphics[height=6cm, width=7cm]{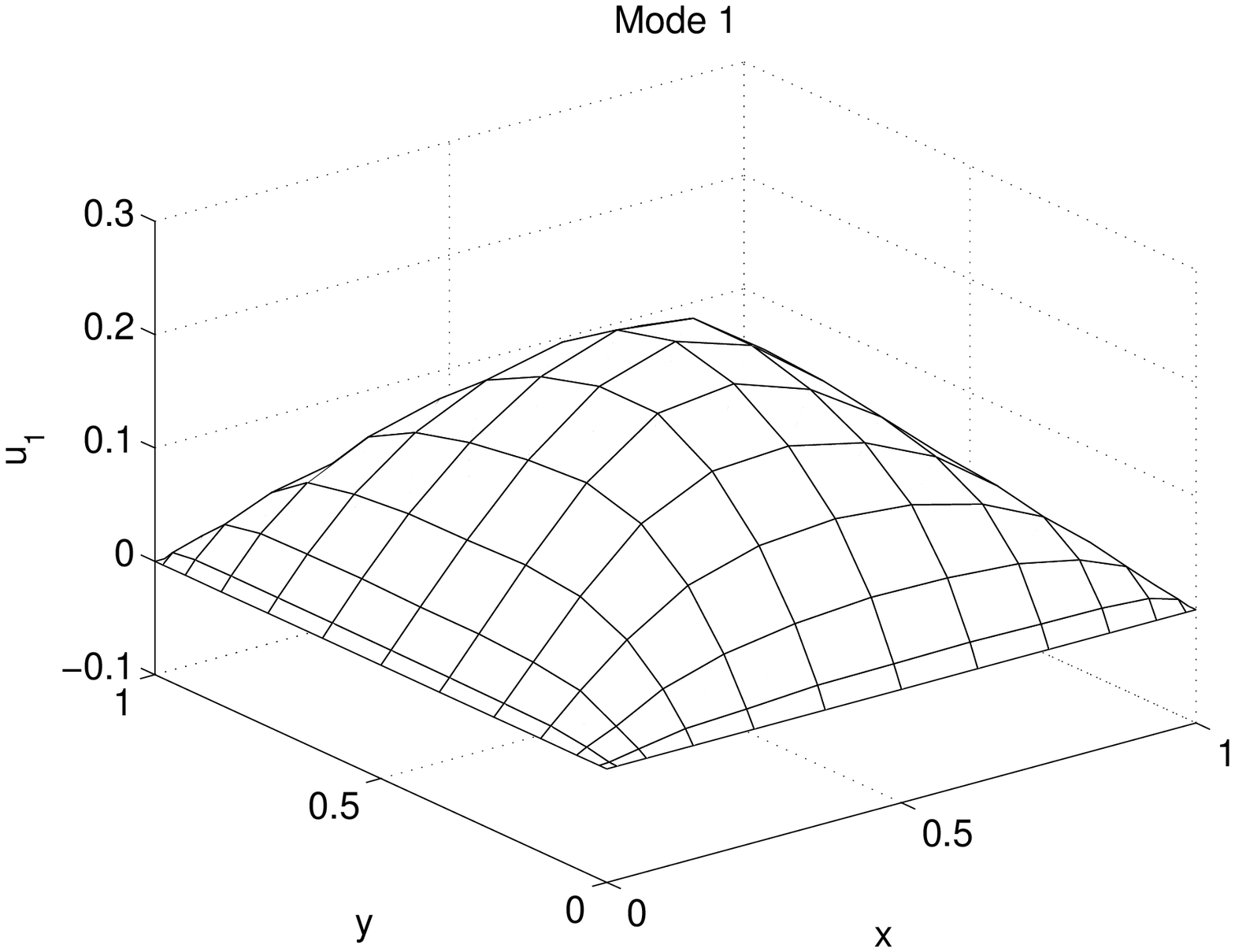}
 \includegraphics[height=6cm, width=7cm]{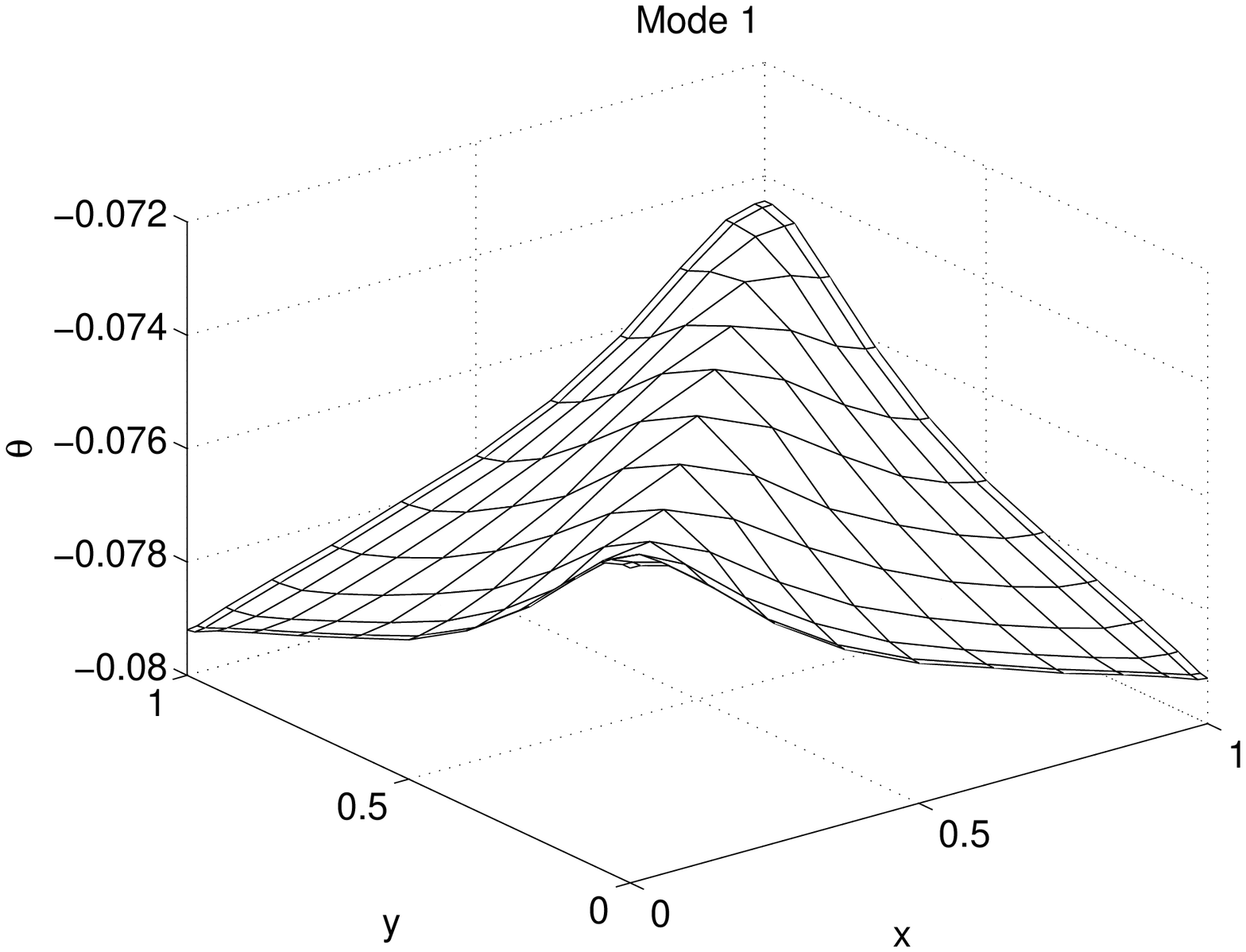} \\
 \includegraphics[height=6cm, width=7cm]{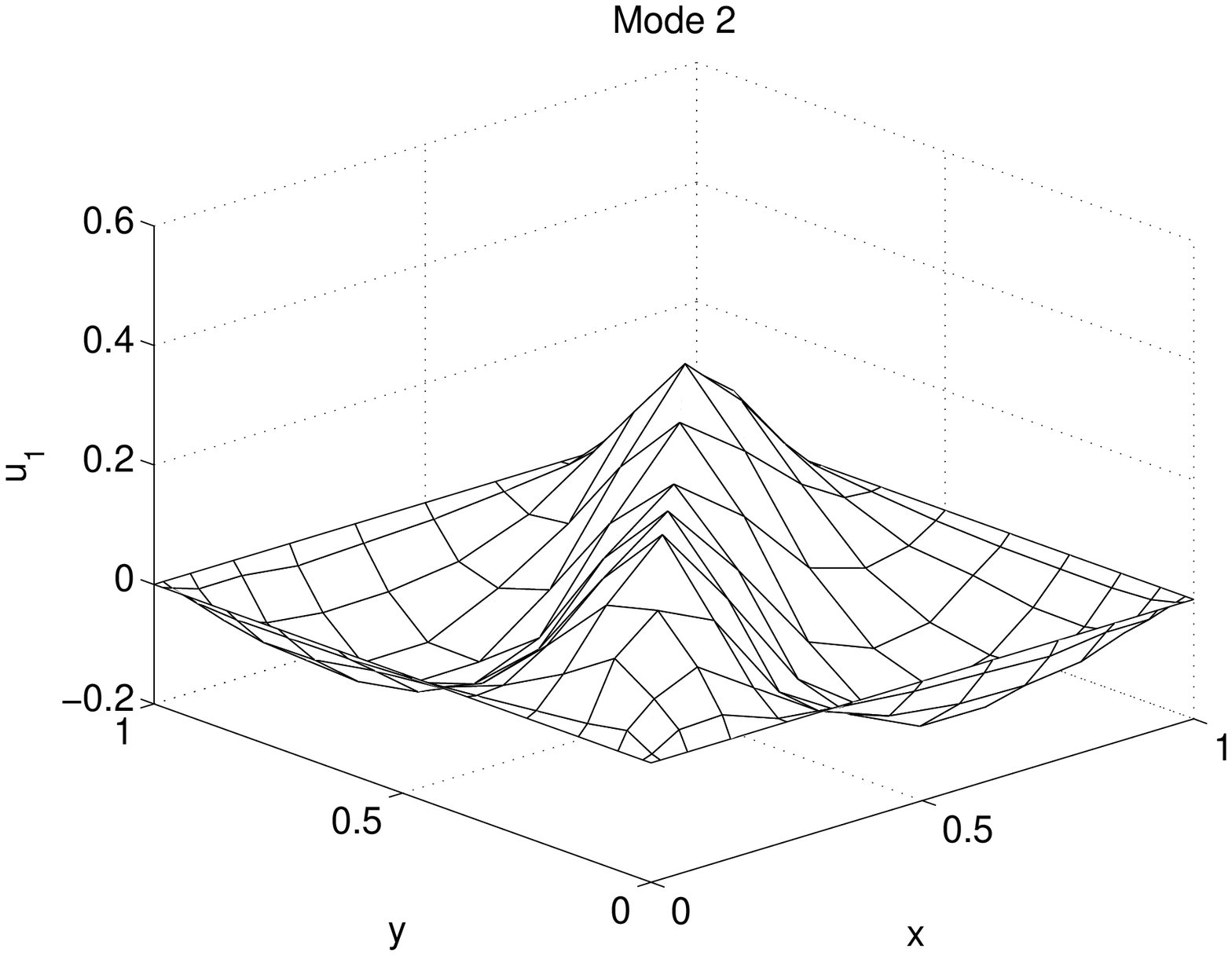}
 \includegraphics[height=6cm, width=7cm]{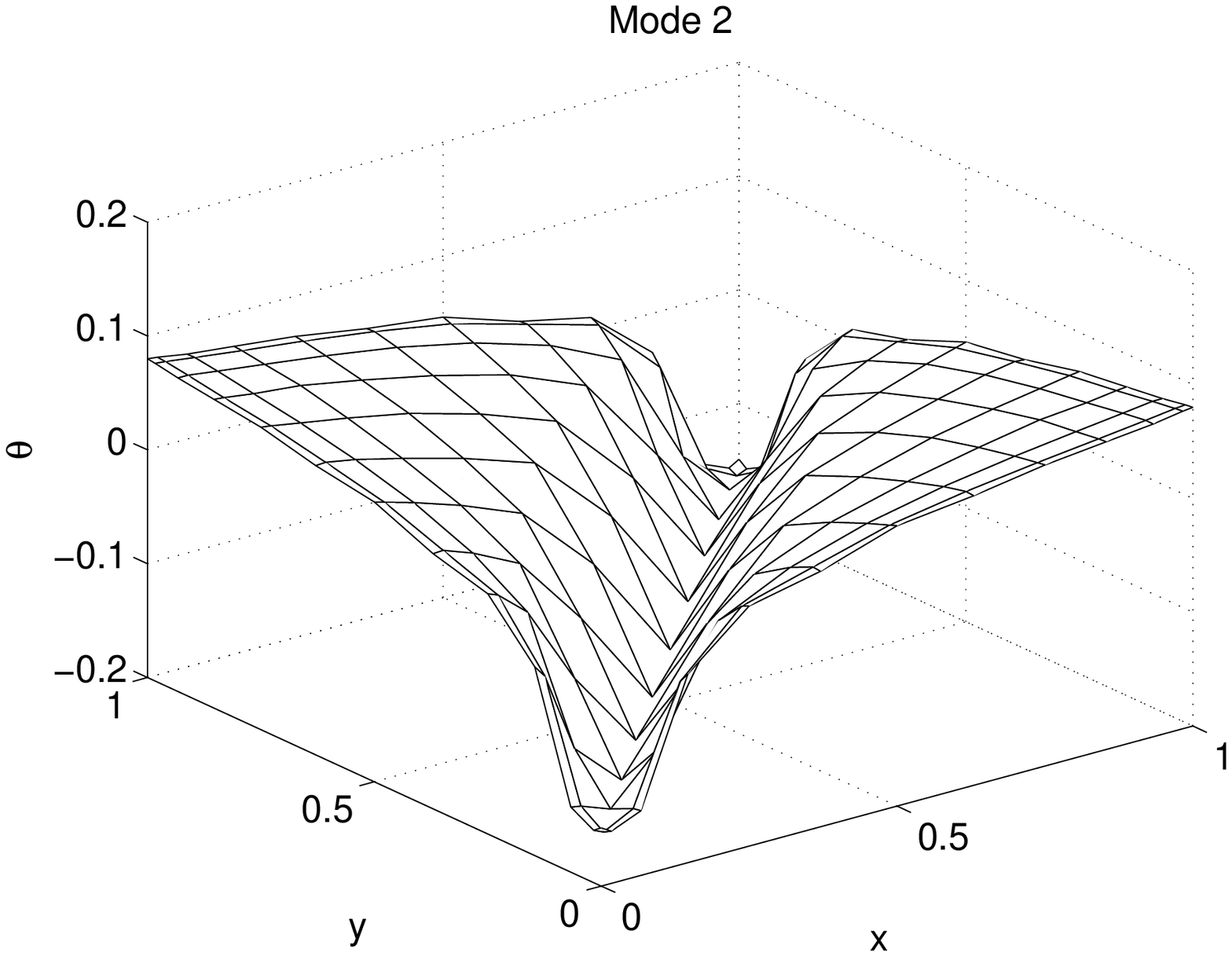} \\
 \includegraphics[height=6cm, width=7cm]{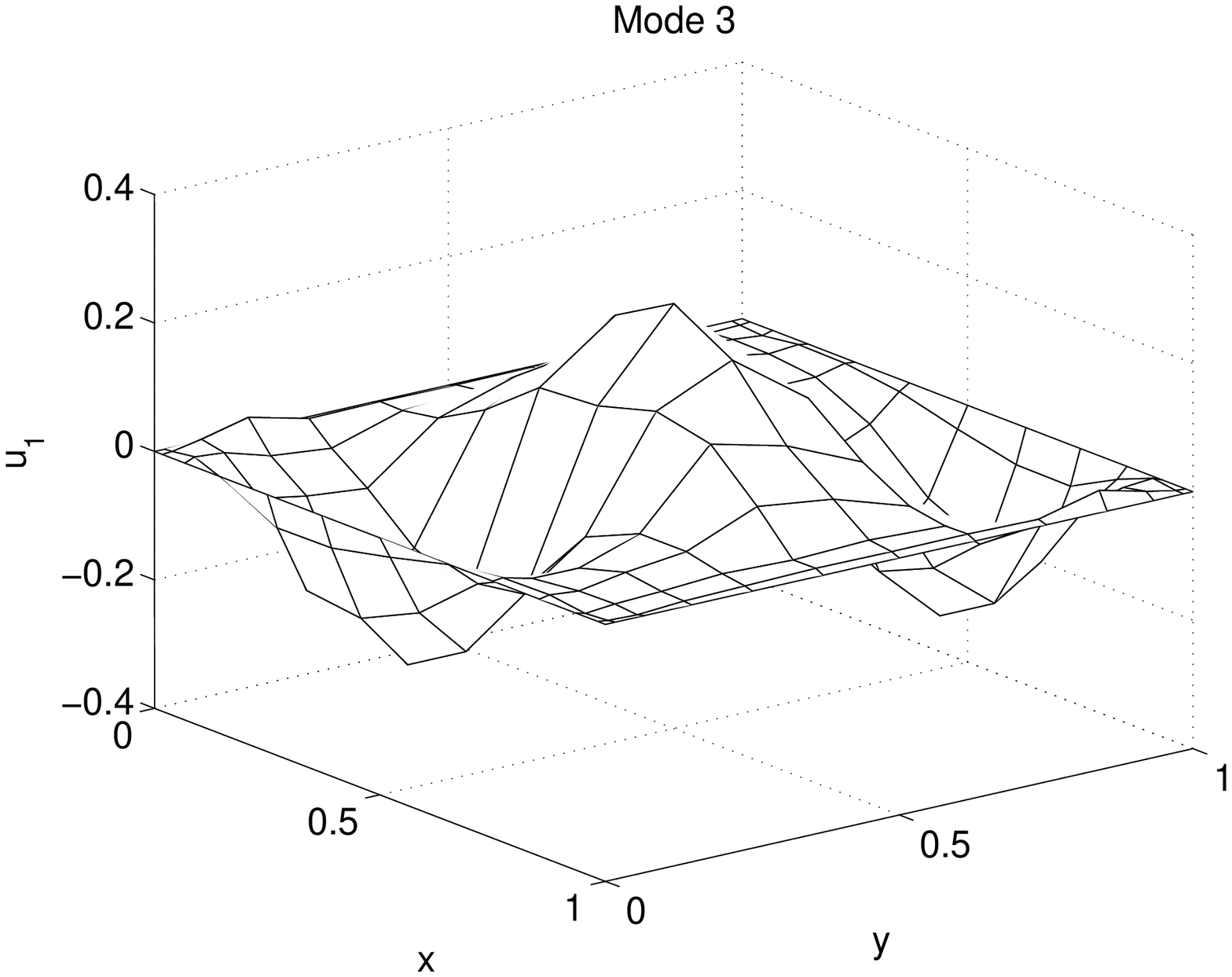}
 \includegraphics[height=6cm, width=7cm]{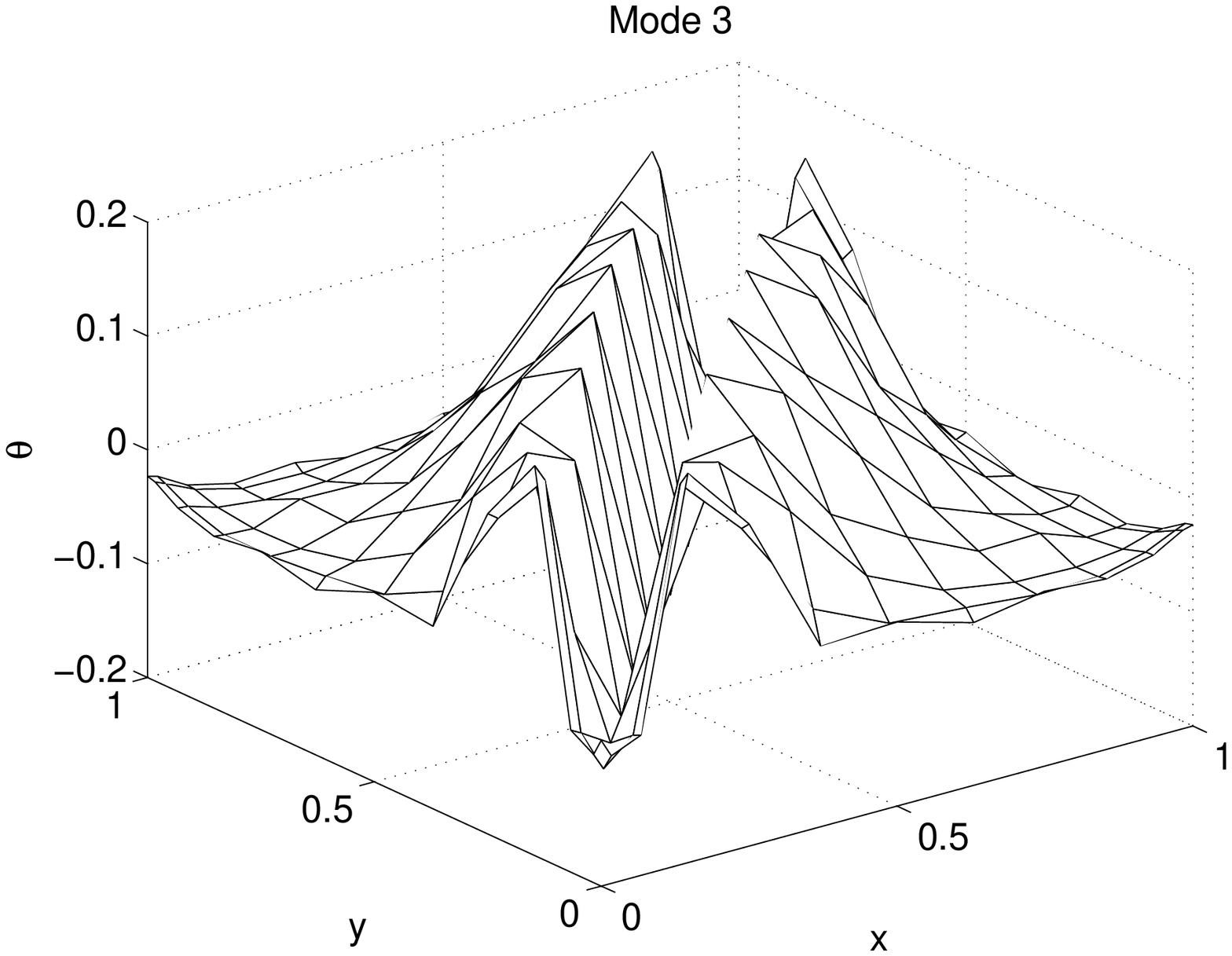} \\
\caption{The first three eigenmodes for $u_1$ and $\theta$ obtained from POD analysis
using collection matrxis constructed by numerical results of PDE model.}
 \label{PODBasis}
  \end{center}         \end{figure}

With the above constructed collection matrix at hand, the POD for each of the field variables can
be done by using Eq.(\ref{podeq-7}) and Eq.(\ref{podeq-8}), by which  the eigenfunctions for
 $u_1, u_2, v_1, v_2$ and
$\theta$ can be estimated \cite{Wang2005}. The first three eigenfunctions for displacement
 $u_1$ are plotted in the left column of Fig.\ref{PODBasis}, together with those
three for temperature in the right column. The eigenfunctions for $v_1$ can be taken as
the same as those for $u_1$ because their differences are in time domain, and similar for $v_2$.

For the construction of the low dimensional model, the first $12$ basis functions
for $u_1, u_2$, and $\theta$ are kept, so the system given by Eq.(\ref{LDSys})
has a dimension of $60$, $12$ for each of the variables $u_1,~u_2,~v_1,~v_2$ and $\theta$ each. A
standard ODE integrator (ode45 in Matlab) is applied to simulate the
state evolution.

\begin{figure}          \begin{center}
 \includegraphics[height=6cm, width=7cm]{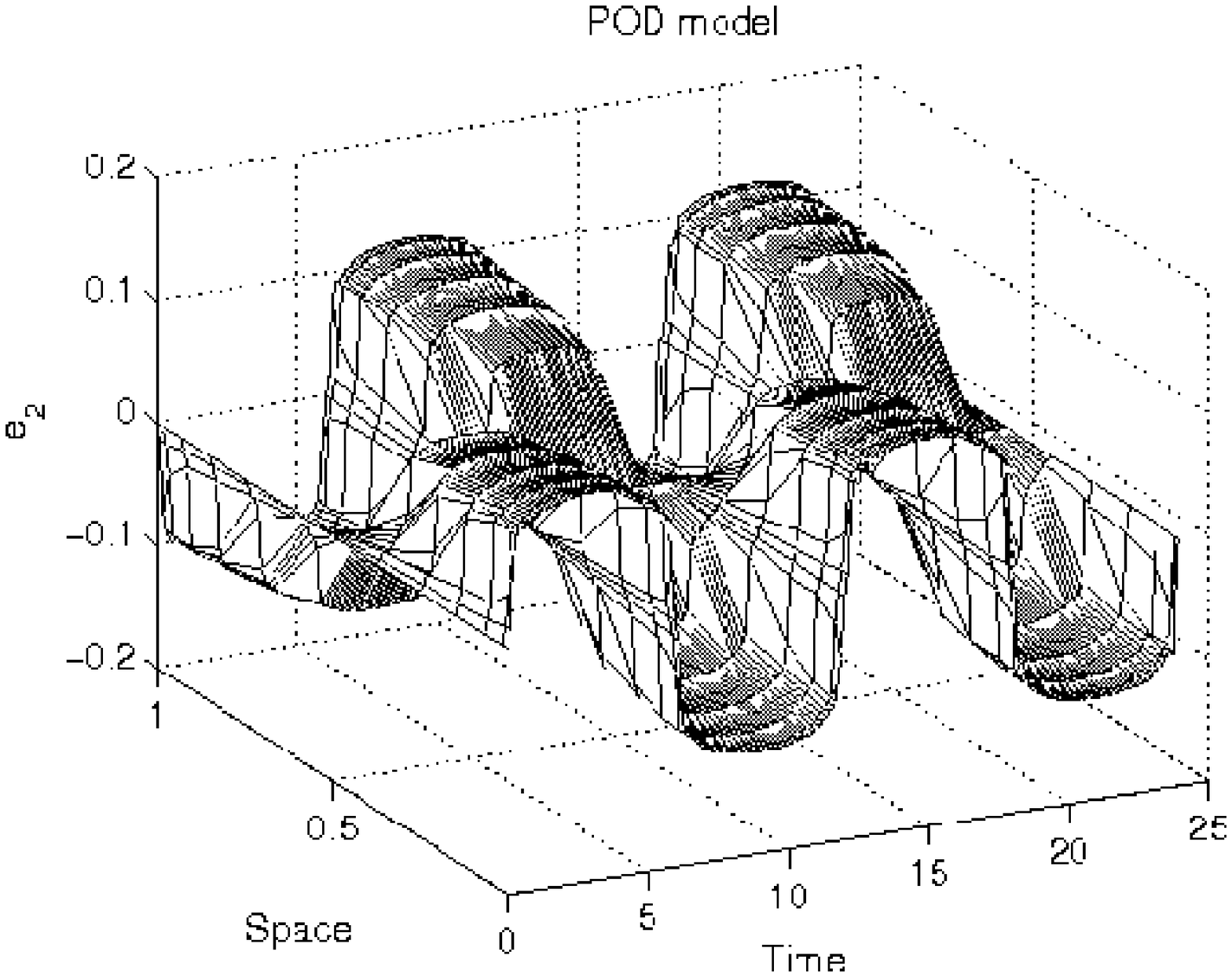}
 \includegraphics[height=6cm, width=7cm]{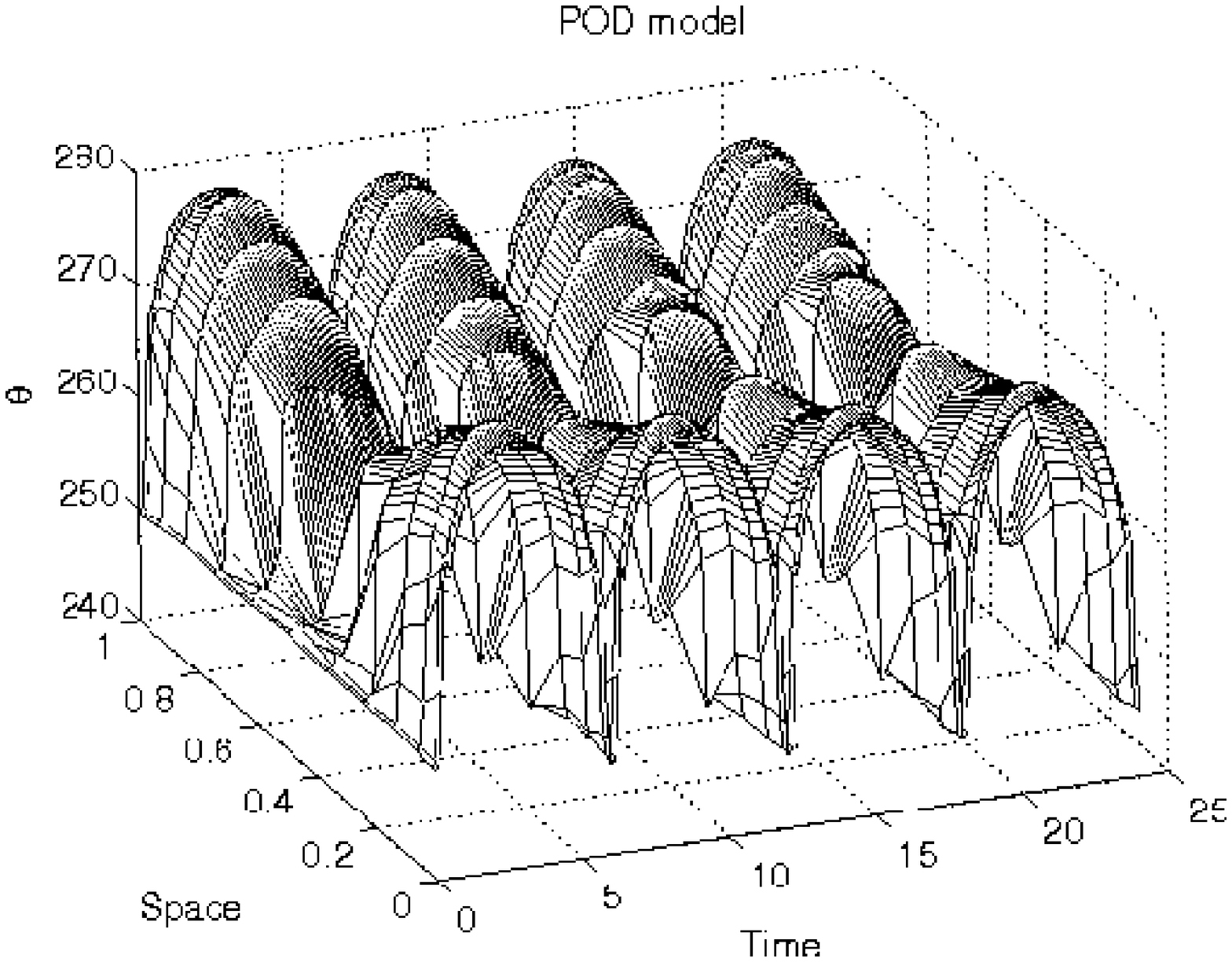}    \\
 \includegraphics[height=6cm, width=7cm]{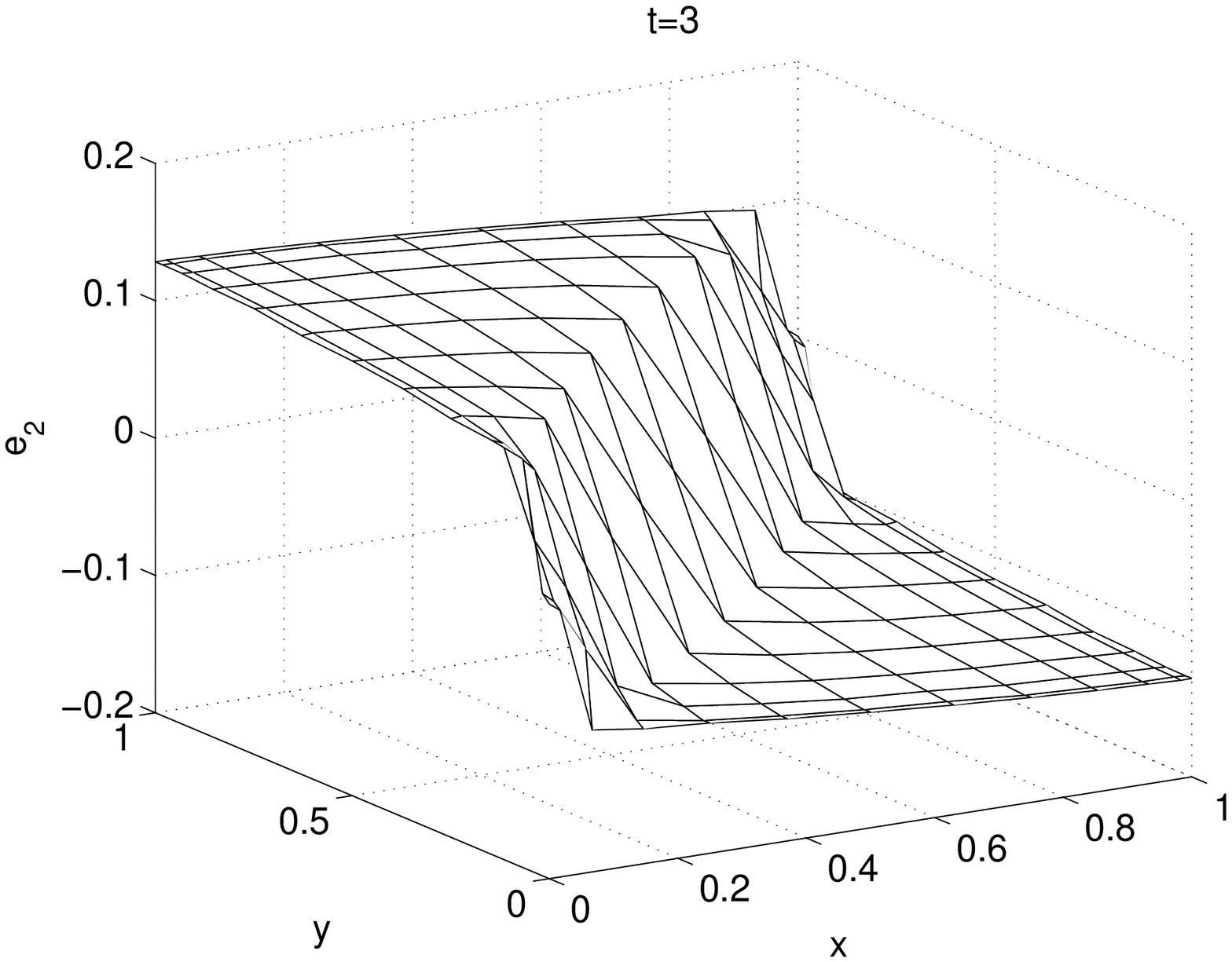}
 \includegraphics[height=6cm, width=7cm]{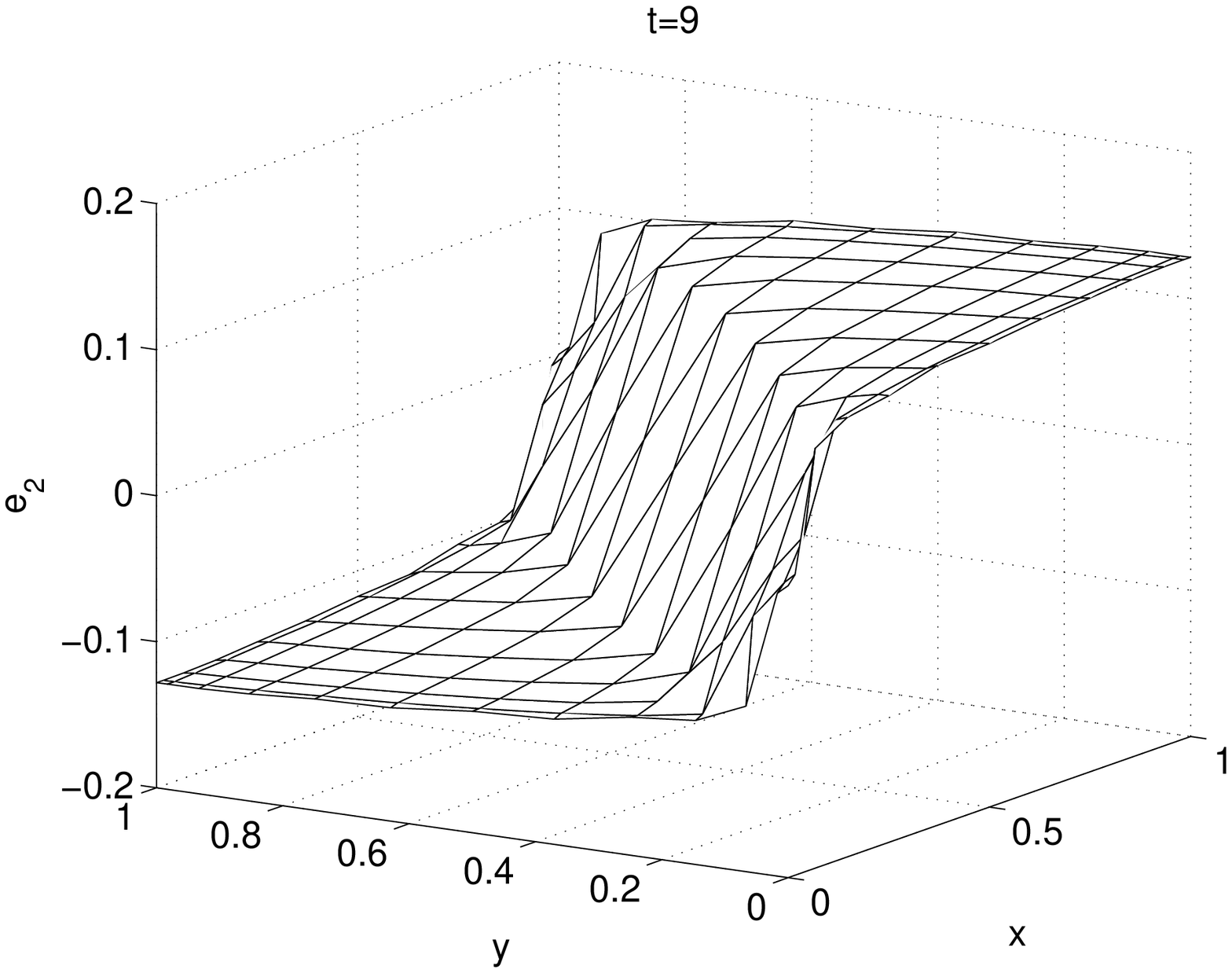} \\
 \includegraphics[height=6cm, width=7cm]{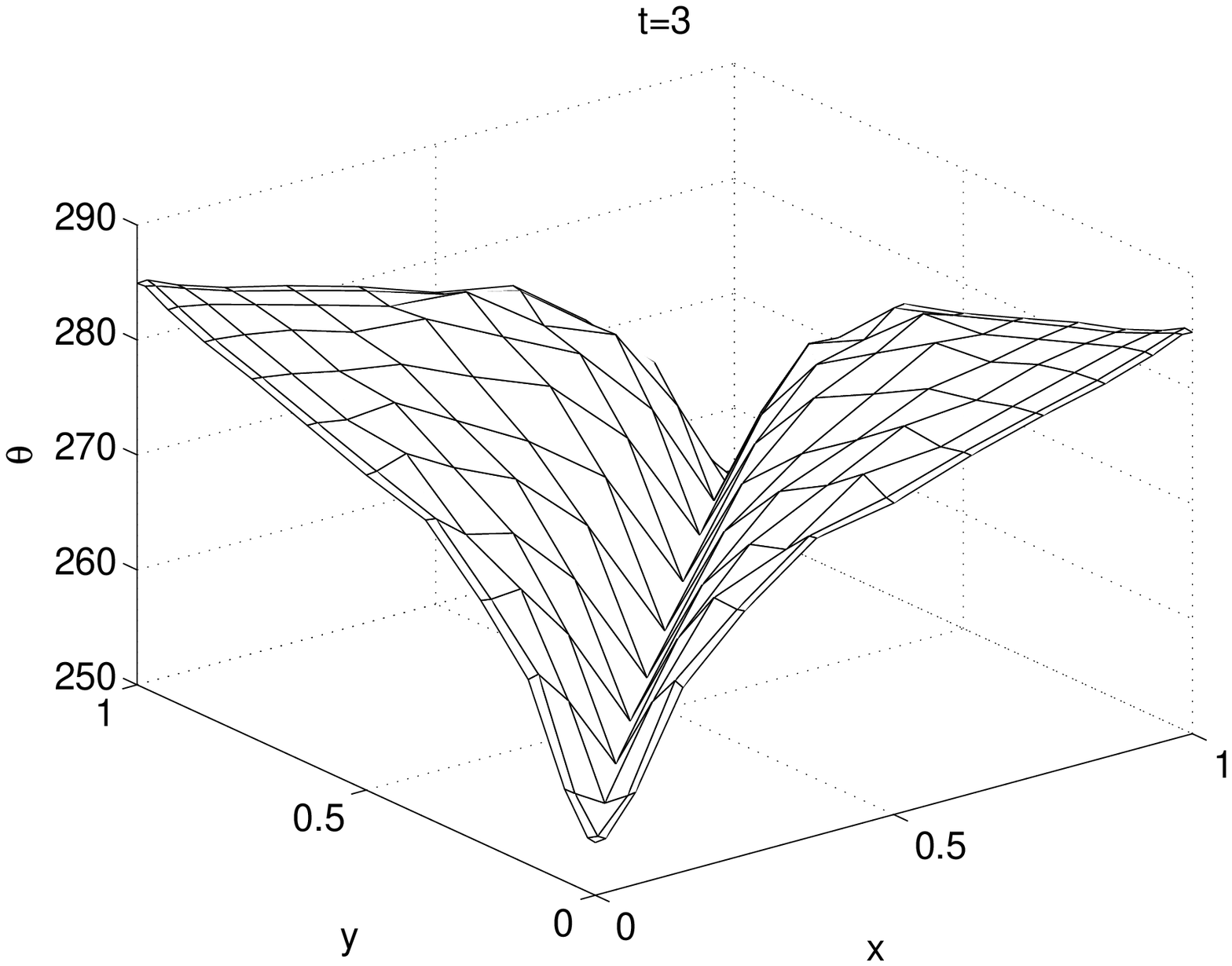}
 \includegraphics[height=6cm, width=7cm]{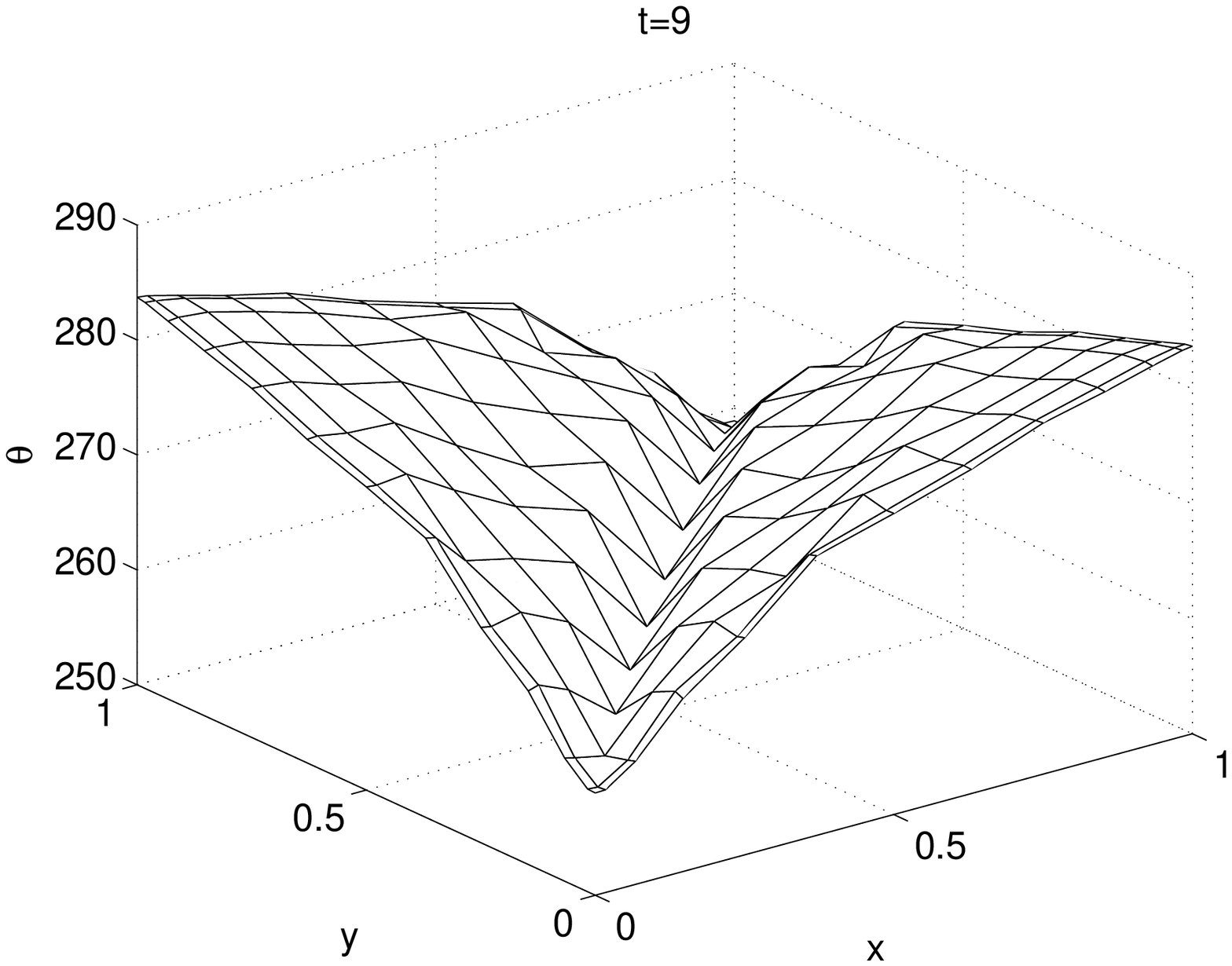}    \\
\caption{Numerical simulation of the dynamics of a shape memory alloy patch using a low
dimensional model. From top to bottom, left to right, (a) Strain evolution on the central horizontal line,
 (b) Temperature evolution on the central horizontal line, (c) Strain distribution at t=3, (d) Strain
distribution at t=9,  (e) Temperature distribution at t=3, (f) Temperature distribution at t=9. }
 \label{PODComp}
  \end{center}         \end{figure}

The mechanical loadings now are $f_x=f_y= 6000 \sin(\pi t /6)$, and the simulation is
performed within the same time interval. The simulated distributions of $e_2$ and
$\theta$ from the low dimensional model are presented in the top row of
Fig.\ref{PODComp}. The simulated distributions clearly indicate the mechanically
induced phase transformation, and the temperature oscillation driven by
the mechanical loads due to the thermomechanical coupling.  To show the effect of
transformation, the distribution of $e_2$ at $t=3$ and  $t=9$ from the low
dimensional model is presented in the middle row of the figure. In this case,  the entire
patch is divided into two sub-areas, associated with the two martensite variants.
The combination of the two martensite variants also depends on the loading
direction \cite{Wang2004}.  To demonstrate the thermo-mechanical coupling, the
temperature distributions $\theta$ at $t=3$ and  $t=9$ also plotted in the figure
(the bottom row).  It can be observed that after the transformation, the
temperature distribution of the patch is also changing.

\begin{figure}          \begin{center}
 \includegraphics[height=6cm, width=7cm]{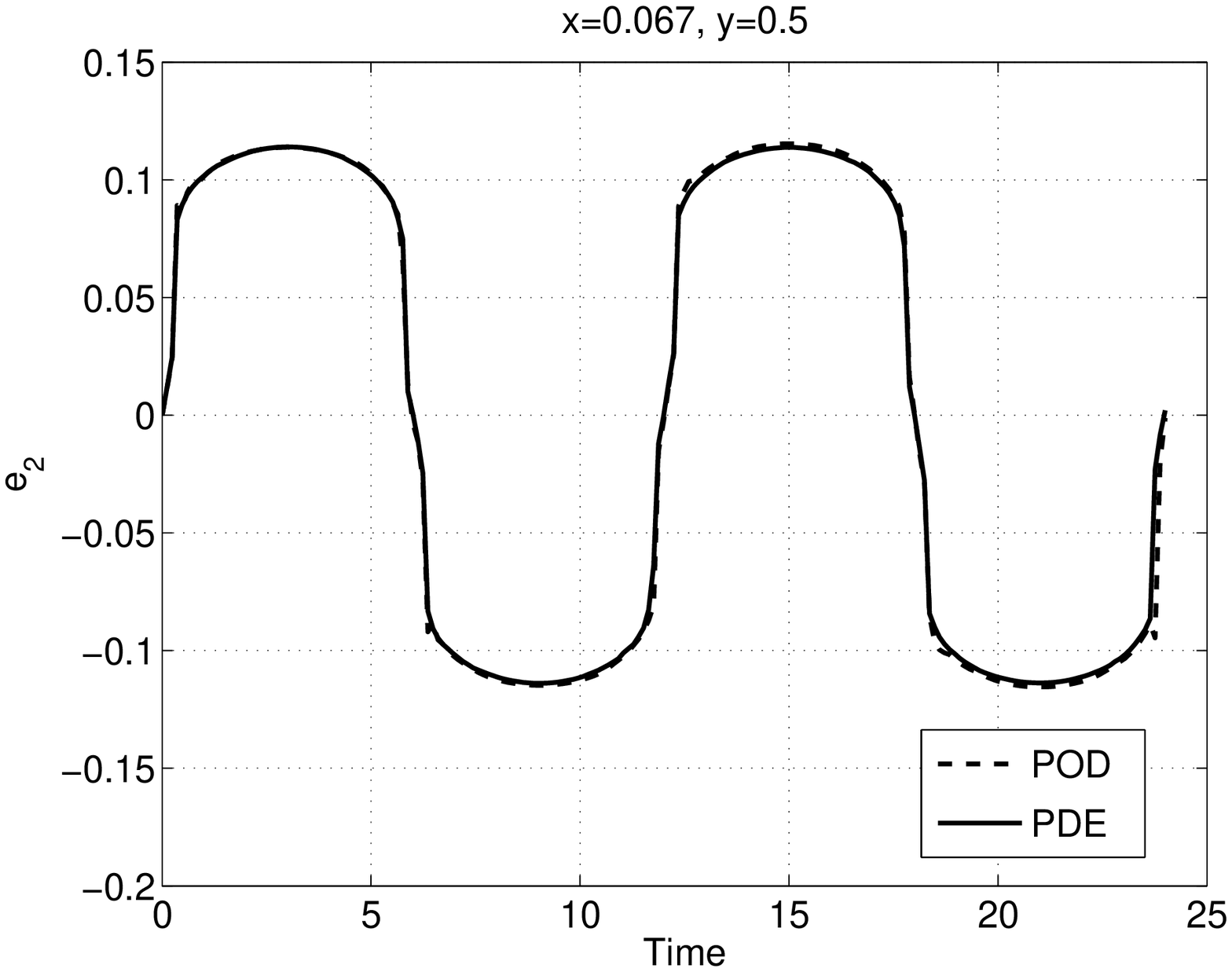}
 \includegraphics[height=6cm, width=7cm]{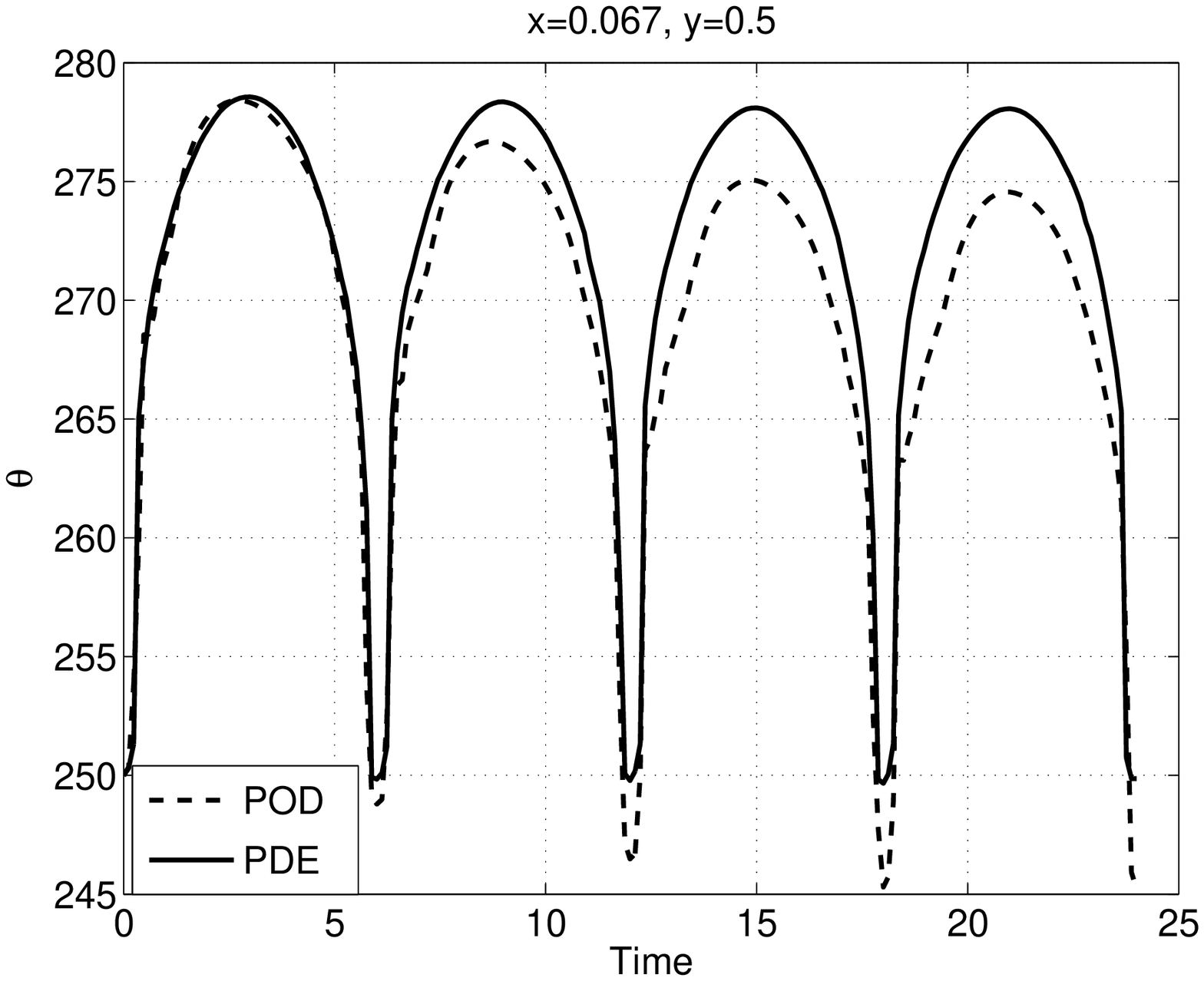}    \\
 \includegraphics[height=6cm, width=7cm]{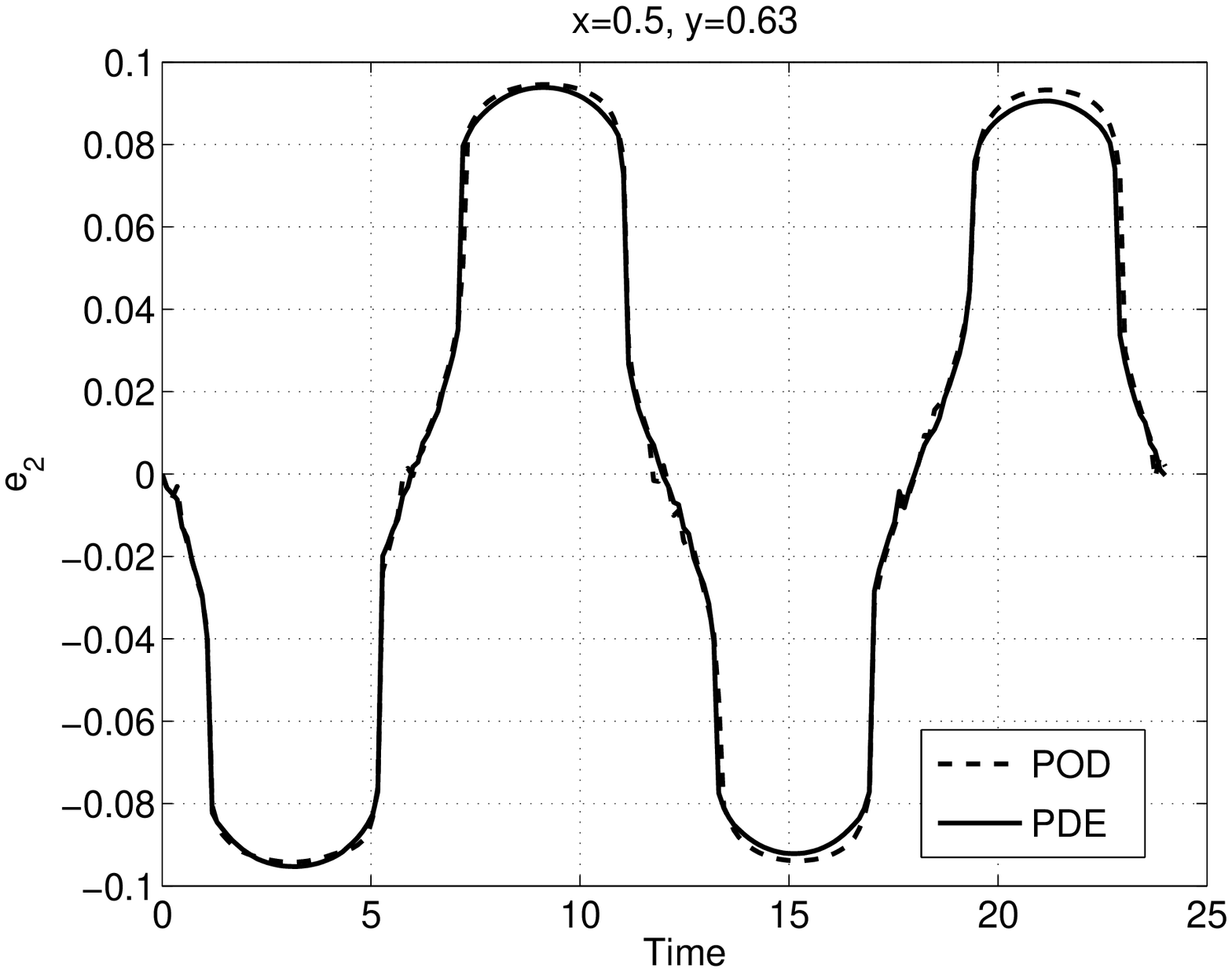}
 \includegraphics[height=6cm, width=7cm]{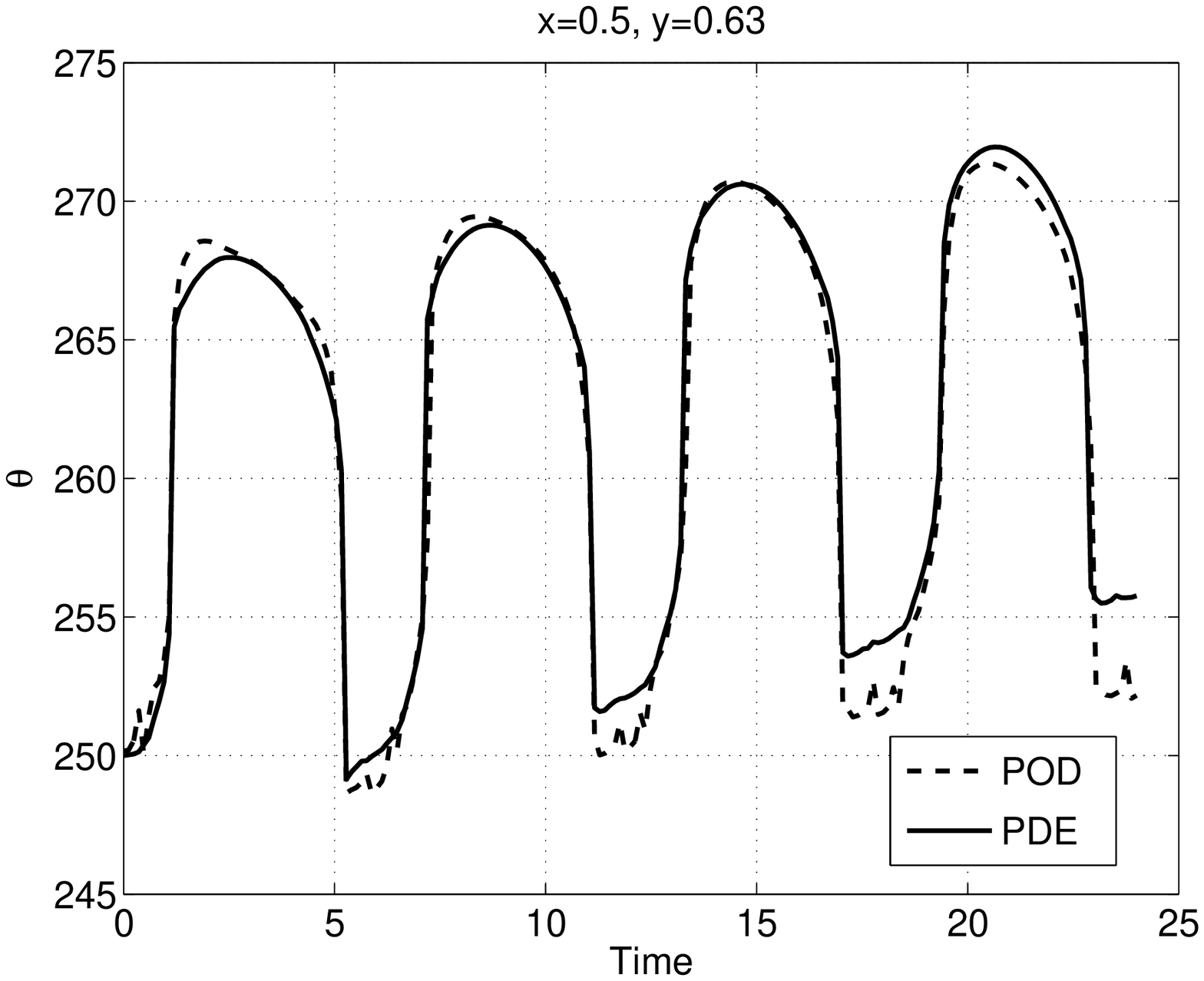} \\
 \includegraphics[height=6cm, width=7cm]{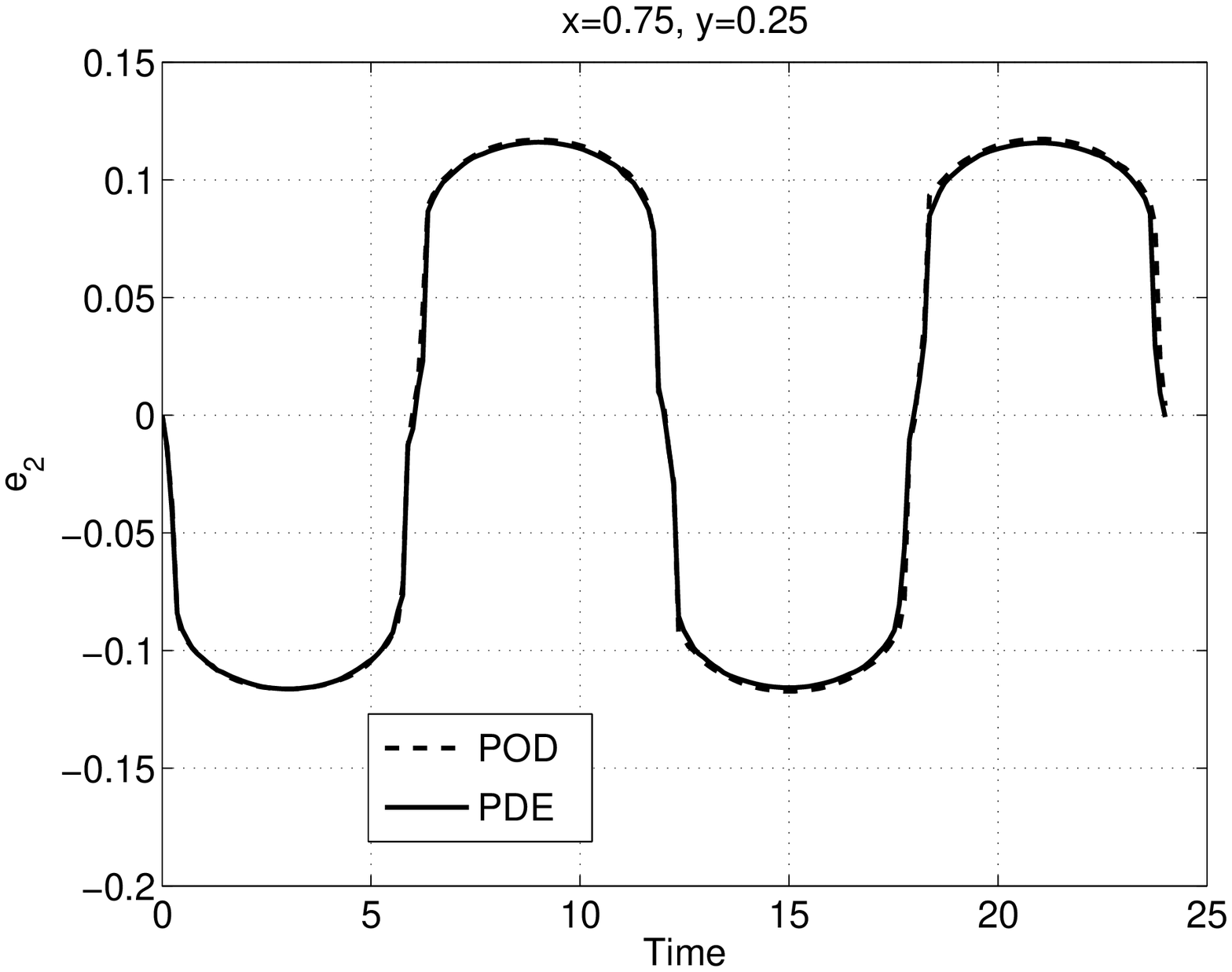}
 \includegraphics[height=6cm, width=7cm]{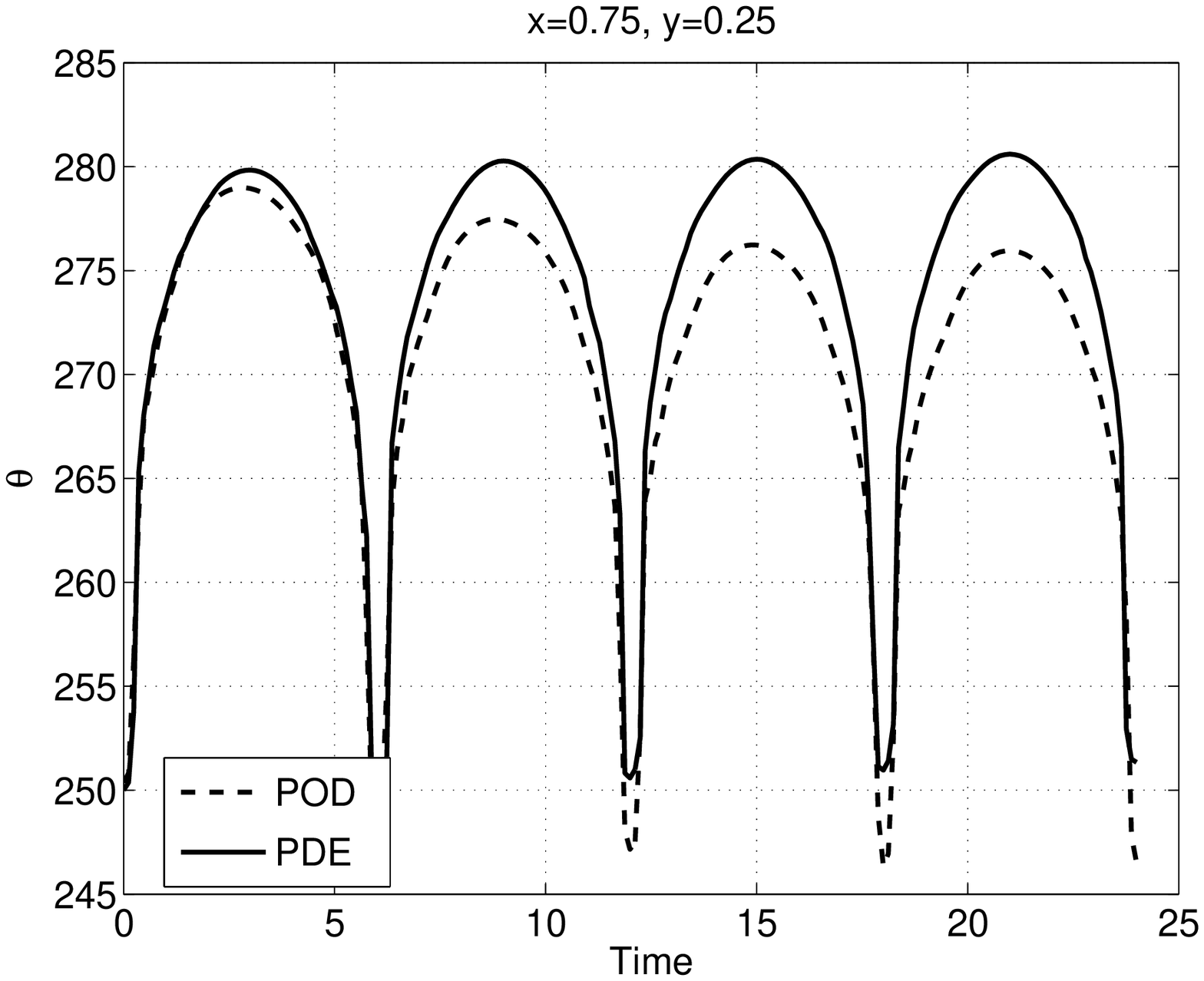} \\
\caption{Comparision of numerical results from the low dimensional model with those
from the PDE model.  From top to bottom, left to right, (a) Strain evolution at (0.067,0.5),
(b) Temperature evolution at (0.067,0.5), (c) Strain evolution at (0.5,0.63),
(d) Temperature evolution at (0.5,0.63), (e) Strain evolution at (0.75,0.25),
(f) Temperature evolution at (0.75,0.25).   }.
 \label{ComparisonTime}
  \end{center}         \end{figure}

To validate the low dimensional model,  a more detailed comparison between the two
sets of  numerical results (from the PDE model and POD model ) is carried out here.
To do so, three points in the shape memory alloy patch are chosen, and the strain
and temperature values from the two models are compared at these points.  The first
point is chosen close to the left boundary as $(0.067, 0.5)$, the strain (left) and temperature
(right) evolutions at the point during the loading process are presented in the top row of
Fig.\ref{ComparisonTime}.  The second point is chosen close to the center of the patch as
 $(0.5, 0.63)$,  and  the strain and temperature evolutions are presented similarly in the middle
row of the figure, while those for the third point $(0.75,0.25)$  are presented in the bottom row.
At all the three points, the strain evolutions simulated using the low dimensional model agree very
well with those from the PDE model, the discrepancies can be distinguished only by a close look at
the plots. The phase transition is indicated clearly by the strain evolution.

The temperature evolutions
from the low dimensional model at the three points also agree qualitatively with their counterparts from
the PDE model,  the oscillatory behaviours of the temperature driven by the mechanical loadings due to
the thermo-mechanical coupling are also successfully captured in the low dimensional model.
The agreement is not perfect in temperature evolutions, there are certain discrepancies between the
two models. This can be explained as  follows.  In the dynamics of the patch, the thermal and mechanical
fields are coupled in the PDE model, but for the low dimensional model it is not practical to extract the
eigenmodes for the thermal and mechanical fields
simultaneously by using the collection matrix including both temperature and displacement distributions,
because their magnitude is very different ( 250  versus 0.1), and characteristics of the variable with smaller
magnitude will be ignored if one do so.  In the current paper,  the eigenmodes for the
displacement distributions are  obtained by the POD analysis using the collection matrix of displacements
only, and similarly for the eigenmodes of temperature distributions. With  this simplification, some
characteristics associated with the coupling between the thermal and mechanical fields are lost.
This loss  will induce minor errors in the low dimensional approximation.  When the dynamics of the
patch are simulated using the low dimensional model with a mechanical loading,  the characteristics of the
thermal behaviour driven by the mechanical loading can be captured, but the minor error due to the partly
loss of coupling characteristics will be accumulated and make the temperature evolutions slightly different
from those of the PDE model, just as indicated by the comparison in Fig.\ref{ComparisonTime}.

From the above numerical experiments and comparison, it is clearly shown that the low dimensional
model is able to capture the characteristics of the dynamics involving phase transformation, as well as
the thermo-mechanical coupling.

\section{Conclusion and Discussion}

In this paper, the dynamics of a ferroelastic patch exhibiting square to
rectangular martensite transformations has been analyzed. The system of PDEs
based on the Landau theory of phase transition has been analyzed numerically
by the Chebyshev
collocation method.  The proper orthogonal decomposition has been  employed
to extract the eigenmodes of the dynamics of the patch.  A low dimension system
has been constructed by projecting the PDEs system onto the subspace spanned by the
first few eigenmodes.

Numerical experiments demonstrated that the complicated dynamics of
ferroelastic materials can be modelled successfully with the derived
low dimensional system.



\newpage

\newpage

\newpage

\newpage

\end{document}